# Optimal Coordination of Flexible DERs in Local Energy and Flexibility Markets to Ensure Social Equity

Niloofar Pourghaderi, Milad Kabirifar, Payman Dehghanian
School of Engineering and Applied Science
George Washington University
Washington D.C., USA

***Abstract*— Local electricity markets offer a promising solution for integrating renewable energy sources and other distributed energy resources (DERs) into distribution networks. These markets enable the effective utilization of flexible resources by facilitating coordination among various agents. Beyond technical and economic considerations, addressing social equity within these local communities is critical and requires dedicated attention in market-clearing frameworks. This paper proposes a social equity-based market-clearing framework for the optimal management of DERs' energy and flexibility within local communities. The proposed framework incorporates consumers' energy burden to ensure fair pricing in energy market clearance. Furthermore, to ensure equity during unbalanced operating conditions, flexible resources are managed in the local flexibility market, ensuring that all participants can trade power fairly under network disturbances. The model is formulated as a second-order cone programming (SOCP) optimization and validated on the IEEE 33-bus test distribution network.**

## NOMENCLATURE

***Indices and sets***

| | |
|---|---|
| $n,m\,/\,l$ | Distribution network nodes/lines. |
| $t$ | Hourly time intervals. |
| $a$ | Local market actors. |
| $p,c,b,e,f\left(d\right)$ | PV units, conventional DGs, BESSs, EVs, and flexible loads (all DERs). |
| $\Omega^{N},\Omega_{n}^{N},\Omega^{N,\mathrm{PCC}}$ | Set of network nodes, connected nodes to node *n*, and substation nodes connected to point of common couplings (PCCs). |
| $\Omega^{T},\Omega^{L}$ | Set of time intervals and network feeders. |
| $\Omega_{n}^{P/C/B/E/F}\left(\Omega_{n}^{D}\right)$ | Set of PV units, conventional DGs, BESS, EVs, and flexible loads (all DERs). |
| $\Omega^{A}\left(\Omega_{n}^{A}\right)$ | Set of actors (actors connected to bus *n*). |

***Parameters***

| | |
|---|---|
| $r_{n,m},x_{n,m}$ | Network lines' resistance and reactance. |
| $\bar{S}_{n,m}^{L}$ | Capacity of distribution network lines. |
| $\underline{V}_{n},\bar{V}_{n}$ | Minimum-maximum voltage magnitude. |
| $\Delta t$ | Duration of time interval. |
| $P_{p,t}^{\mathrm{PV}},PF_{p,t}^{\mathrm{PV}},S_{p}^{\mathrm{PV}}$ | PVs' forecasted power, power factor, and capacity. |
| $\bar{P}_{b/e}^{\text{ch-ESS/EV}},\bar{P}_{b/e}^{\text{dch-ESS/EV}}$ | BESS/EV charge-discharge power limits. |
| $\eta_{b}^{\text{ch-ESS/EV}},\eta_{b}^{\text{dch-ESS/EV}}$ | BESS/EV charge-discharge efficiencies. |
| $E_{b}^{\text{0-ESS}},\bar{S}_{b}^{\mathrm{ESS}}$ | Initial charge and capacity of BESSs. |
| $\underline{SoC}_{b/e}^{\mathrm{ESS/EV}},\overline{SoC}_{b/e}^{\mathrm{ESS/EV}}$ | BESS/EV SoC limits. |
| $a_{e}^{\mathrm{EV}},d_{e}^{\mathrm{EV}}$ | Arrival and departure times of EVs. |
| $\lambda_{t}^{\mathrm{UG}}$ | Energy price of upstream grid. |
| $P_{n,t}^{\mathrm{Fixed}}$ | Customers' inflexible load. |
| $\mathrm{EB}_{a,t},\mathrm{EB}_{n,t}$ | Energy burden of actors and load nodes. |
| $C_{c}^{\mathrm{DG}},C_{f}^{\mathrm{FL}},C_{b}^{\mathrm{ESS}},C_{e}^{\mathrm{EV}}$ | DGs, flexible loads, BESSs, and EVs' operating costs. |
| $\underline{P}_{c}^{\mathrm{DG}},\bar{P}_{c}^{\mathrm{DG}},RR_{c}^{\text{D/U-DG}}$ | DGs' power and ramp-rate limits. |
| $\bar{P}_{f,t}^{\mathrm{FL}},\underline{E}_{n}^{\mathrm{L}}$ | Flexible load power, load energy limits. |

***Variables***

| | |
|---|---|
| $\upsilon_{n,t},\ell_{n,m,t}$ | Square of bus voltage, and line current. |
| $p_{n,m,t}^{\mathrm{F}},q_{n,m,t}^{\mathrm{F}}$ | Lines' active and reactive power flow. |
| $p_{n,t}^{\mathrm{G}},p_{n,t}^{\mathrm{L}}$ | Net generation and consumption power. |
| $p_{t}^{\mathrm{UG}}$ | Upstream grid active power. |
| $\lambda_{n,t}^{\mathrm{E}}$ | DLMP of distribution network nodes. |
| $p_{p,t}^{\mathrm{PV}},p_{c,t}^{\mathrm{DG}},p_{f,t}^{\mathrm{FL}}$ | PVs, DGs and flexible loads power. |
| $x_{b/e,t}^{\text{ch-ESS/EV}},x_{b/e,t}^{\text{dch-ESS/EV}}$ | BESS/EV charge-discharge indicators. |
| $p_{b/e,t}^{\text{ch-ESS/EV}},p_{b/e,t}^{\text{dch-ESS/EV}}$ | BESS/EV Charging-discharging power. |
| $SoC_{b/e,t}^{\mathrm{ESS/EV}}$ | BESS/EV state of charge. |
| $UF_{d,t}^{D},DF_{d,t}^{D}$ | DERs' upward-downward flexibility. |
| $y_{d,t}^{\mathrm{UF},D},y_{d,t}^{\mathrm{DF},D}$ | Upward-downward flexibility binary indicators of DERs. |

## I. INTRODUCTION

Energy equity has become a critical criterion, gaining increasing attention from policymakers, particularly in the context of the green energy transition [1]. As the closest segment of the power grid to end-users, distribution networks play a pivotal role in promoting social equity. Leveraging the potential of distributed energy resources (DERs) within these networks can help ensure fair energy distribution among consumers. Moreover, it is essential that DERs are afforded the opportunity to participate in energy trading in a just and equitable manner. Local electricity markets provide an effective platform to coordinate the strategies of consumers, producers, and prosumers dispersed throughout the distribution network, enabling optimal utilization of their resources [2]. The flexibility offered by distributed resources within these networks holds significant promise for fostering equitable power trading among all participants in local communities. This paper focuses on optimal utilization of this potential to ensure fairness and equity in local markets.

The coordination of DERs through local electricity market frameworks has been extensively explored in the literature. In [3] and [4], the technical and economic aspects of local electricity markets are examined. Furthermore, references [5] and [6] developed multi-agent frameworks to address the roles of various stakeholders within these local communities. While these studies focus on the technical and economic dimensions, the social aspects of local communities, particularly concerning energy equity, remain underexplored in local electricity market frameworks.

Energy equity has been investigated in several areas of power systems. For instance, [1] and [7] discuss the impact of low-carbon energy transition policies on social equity, while [8] explores equity implications in the shift toward decentralized energy systems. Concerns about energy equity within current electric system designs are addressed in [9] and [10]. Furthermore, [11] investigates the influence of local energy generators on equity and distributional justice, highlighting associated challenges. These studies primarily focus on equity from policy perspectives and conceptual clarifications. Reference [12] develops an equity-aware restoration model for power distribution networks, ensuring that after power outages, loads are restored equitably. In [13], mobile charging stations are employed as temporary charging solutions to provide socially equitable access for EV owners to smart parking lots. The research utilizes prioritized demand indices to optimize the locations of these charging stations. Authors in [14] incorporate energy equity into wholesale energy market clearing through a multi-layer model, where consumers are categorized based on an equity index, and the market is cleared separately for each class.

Based on the reviewed references, the literature lacks the model that incorporate equity criteria into the optimal coordination of DERs within local electricity market frameworks. Moreover, to the best of the authors' knowledge, this study is the first to leverage the flexibility of various energy resources through a local market framework to ensure social equity for all participants under both normal and unbalanced network operating conditions. This paper proposes a two-stage framework to address social equity in the clearing of local energy and flexibility markets. In the first stage, the optimal energy management of DERs is achieved by incorporating consumers' energy burden into the pricing mechanism. In the second stage, the flexibility of diverse DERs in the power distribution network is utilized through a local flexibility market to uphold equity within the community during network disturbances.

## II. EQUITY-BASED LOCAL MARKET FRAMEWORK

### A. General Model

This paper presents a clearing framework for day-ahead local energy and real-time local flexibility markets, addressing energy equity alongside technical and economic considerations. As illustrated in Fig. 1, the local market operator (which could be a distribution system operator (DSO)) is responsible for clearing energy and flexibility markets. The DSO is a technical agent which aims to minimize operational costs and incorporates the operational constraints of the underlying distribution network into the market-clearing process. Other participants in the local community include consumers with fixed and flexible loads, electric vehicles (EVs), prosumers equipped with distributed generators (DGs) (e.g., rooftop photovoltaic (PV) panels and conventional DGs), battery energy storage systems (BESSs), and producers with DGs and BESSs.

In the first stage, the local energy market is cleared with the objective of minimizing total costs, using a distribution locational marginal pricing (DLMP) mechanism. DLMP reflects the energy price at each bus of the distribution network [15]. The DLMP reflects the economic and technical aspects of the network and DERs; however, the social aspect is not addressed. As a result, customers of all income levels are treated the same, which does not align with social equity principles. To ensure energy equity among local community participants, an energy burden index is incorporated to adjust DLMP signals for each participant. The energy burden index, defined as the ratio of energy costs to total income, is particularly significant for low-income consumers, who typically have higher energy burdens [16]. By optimizing DER energy management in the local energy market, the model determines the optimal energy dispatch for consumers, prosumers, and producers, which is subsequently passed to the second stage. The second stage leverages the flexibility of DERs to maintain energy equity during disturbances. Disturbances include sudden changes in consumption or generation. In this stage, the local flexibility market is settled in real time to minimize load curtailment and ensure that any necessary load shedding is distributed equitably among participants. Flexible resources are adjusted within the local flexibility market to ensure network balance and fair operation under such abnormal conditions.

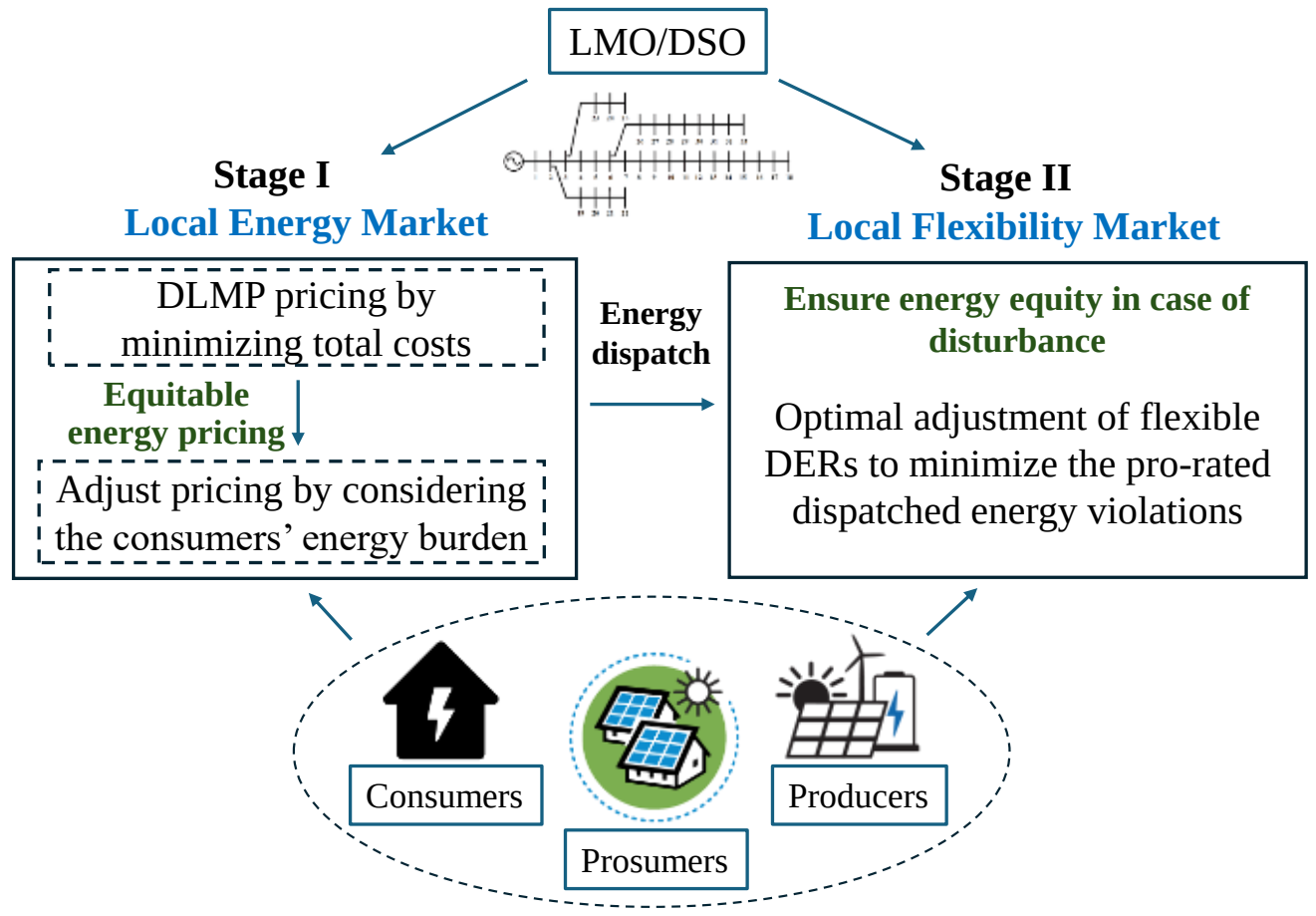

Figure 1. General framework for clearing local energy and flexibility markets by ensuring energy equity.

The proposed two-stage model is formulated as a second-order cone programming (SOCP) optimization problem. This formulation efficiently captures the impact of losses and line congestion on DLMP pricing and guarantees a globally optimal solution within a reasonable computational time.

## B. Mathematical Formulation

### 1) Stage I: Local Energy Market

The objective of the local energy market clearing is to minimize the total operational costs of the system, as expressed in (1). Following the market clearing, the DLMP signals are adjusted to account for the energy burden index, ensuring equitable electricity pricing for consumers within the local community.

$$\textbf{Min} \sum_{t\in\Omega^T}\left( p_t^{\mathrm{UG}}\lambda_t^{\mathrm{UG}} + \sum_{n\in\Omega^N}\left( \begin{array}{l} \sum_{c\in\Omega_n^C} p_{c,t}^{\mathrm{DG}} C_c^{\mathrm{DG}} + \sum_{f\in\Omega_n^F} p_{f,t}^{\mathrm{FL}} C_f^{\mathrm{FL}} \\ \sum_{b\in\Omega_n^B}\left(p_{b,t}^{\text{ch-ESS}} + p_{b,t}^{\text{dch-ESS}}\right) C_b^{\mathrm{ESS}} \\ + \sum_{e\in\Omega_n^E}\left(p_{e,t}^{\text{ch-EV}} + p_{e,t}^{\text{dch-EV}}\right) C_e^{\mathrm{EV}} \end{array}\right)\right) \quad (1)$$

In the proposed mathematical model, the generating power of DGs, the charging/discharging power of BESSs and EVs, and the consumption power of both fixed and flexible loads are aggregated at each bus. The net generation and consumption power at each bus are calculated as follows:

$$p_{n,t}^{\mathrm{G}} = \sum_{c\in\Omega_n^C} p_{c,t}^{\mathrm{DG}} + \sum_{p\in\Omega_n^P} p_{p,t}^{\mathrm{PV}} + \sum_{b\in\Omega_n^B} p_{b,t}^{\text{dch-ESS}} + \sum_{e\in\Omega_n^E} p_{e,t}^{\text{dch-EV}}, \quad \forall n\in\Omega^N, t\in\Omega^T \quad (2)$$

$$p_{n,t}^{\mathrm{L}} = P_{n,t}^{\mathrm{Fixed}} + \sum_{f\in\Omega_n^F} p_{f,t}^{\mathrm{FL}} + \sum_{b\in\Omega_n^B} p_{b,t}^{\text{ch-ESS}} + \sum_{e\in\Omega_n^E} p_{e,t}^{\text{ch-EV}}, \quad \forall n\in\Omega^N, t\in\Omega^T \quad (3)$$

The active and reactive power balance must be satisfied at each bus. The set of constraints in Equation (4) addresses the active power balance, where $\lambda_{n,t}^{\mathrm{E}}$ is the corresponding dual variable that represents the DLMP for the local energy market.

$$\sum_{m\in\Omega_n^N} p_{n,m,t}^{\mathrm{F}} + p_{n,t}^{\mathrm{L}} = p_{n,t}^{\mathrm{G}} + p_t^{\mathrm{UG}}\Big|_{n\in\Omega^{N,\mathrm{PCC}}} : \lambda_{n,t}^{\mathrm{E}}, \forall n\in\Omega^N, t\in\Omega^T \quad (4)$$

To capture the effects of network losses, line congestion, and voltage support in the DLMP calculations, the SOCP model for the distribution network load flow is adopted [17]. This model is a relaxed form of semi-definite programming, which transforms the problem into a convex optimization problem. The associated distribution network constraints are represented in Equations (5) – (8).

$$\upsilon_{n,t} - \upsilon_{m,t} - 2\left(r_{n,m}.p_{n,m,t}^{\mathrm{F}} + x_{n,m}.q_{n,m,t}^{\mathrm{F}}\right) + \ell_{n,m,t}\left(r_{n,m}^2 + x_{n,m}^2\right) = 0 \quad \forall n\in\Omega^N, m\in\Omega_n^N, t\in\Omega^T \quad (5)$$

$$\left(p_{n,m,t}^{\mathrm{F}}\right)^2 + \left(q_{n,m,t}^{\mathrm{F}}\right)^2 \le \upsilon_{n,t}\ell_{n,m,t}, \quad \forall n\in\Omega^N, m\in\Omega_n^N, t\in\Omega^T \quad (6)$$

$$\left(p_{n,m,t}^{\mathrm{F}}\right)^2 + \left(q_{n,m,t}^{\mathrm{F}}\right)^2 \le \left(\bar{S}_{n,m}^{L}\right)^2, \quad \forall n\in\Omega^N, m\in\Omega_n^N, t\in\Omega^T \quad (7)$$

$$\left(\underline{V}_n\right)^2 \le \upsilon_{n,t} \le \left(\bar{V}_n\right)^2, \quad \forall n\in\Omega^N, t\in\Omega^T \quad (8)$$

To ensure a fair distribution of energy prices among consumers, the energy burden index is used to adjust the DLMP for customers located at each bus in the distribution network. Initially, the DLMP at each bus is adjusted based on the average energy burden of the customers served by that bus. Subsequently, the energy price for each customer is adjusted based on their individual energy burden and the new DLMP of the corresponding bus (Equations (9) – (11)).

$$\lambda_{n,t}^{\text{E-new}} = \lambda_{n,t}^{\mathrm{E}}\left(\mathrm{EB}_{n,t} / \tilde{\mathrm{EB}}_t\right), \forall n\in\Omega^N, t\in\Omega^T \quad (9)$$

$$\lambda_{a,t}^{\text{E-new}} = \lambda_{n,t}^{\text{E-new}}\left(\mathrm{EB}_{a,t} / \mathrm{EB}_{n,t}\right), \quad \forall a\in\Omega_n^A, t\in\Omega^T \quad (10)$$

$$\sum_{a\in\Omega_n^A} \lambda_{a,t}^{\text{E-new}}.p_{a,t}^{\mathrm{L}} = \lambda_{n,t}^{\text{E-new}}.\sum_{a\in\Omega_n^A} p_{a,t}^{\mathrm{L}}, \forall n\in\Omega^N, t\in\Omega^T \quad (11)$$

The model of the actors' resources in the local energy market is described as follows. The power generated by PV units is constrained by both the minimum forecasted generation capability and the capacity of the PV unit's inverter, as expressed in Equation (12). The apparent power generated by a PV unit is subject to certain limitations which is stated in (13).

$$p_{p,t}^{\mathrm{PV}} \le \mathrm{Min}\left\{P_{p,t}^{PV}, PF_{p,t}^{\mathrm{PV}}.S_p^{\mathrm{PV}}\right\}, \quad \forall p\in\Omega_n^P, t\in\Omega^T \quad (12)$$

$$\left(p_{p,t}^{\mathrm{PV}}\right)^2 + \left(q_{p,t}^{\mathrm{PV}}\right)^2 \le \left(S_p^{\mathrm{PV}}\right)^2, \quad \forall p\in\Omega_n^P, t\in\Omega^T \quad (13)$$

To model the conventional DGs, their output power must adhere to the limits based on (14), and the ramp rate constraint, as defined in (15), must also be satisfied.

$$\underline{P}_c^{\mathrm{DG}} \le p_{c,t}^{\mathrm{DG}} \le \bar{P}_c^{\mathrm{DG}}, \quad \forall c\in\Omega_n^C, t\in\Omega^T \quad (14)$$

$$-RR_c^{\text{D-DG}}.\Delta t \le p_{c,t}^{\mathrm{DG}} - p_{c,t-1}^{\mathrm{DG}} \le RR_c^{\text{U-DG}}.\Delta t, \forall c\in\Omega_n^C, t\in\Omega^T \quad (15)$$

Actors' EVs are modeled by (16) – (22). The charging and discharging operations of EVs are represented by (16) – (18), while the EV's SoC is modeled by (19) – (22). Constraint (22) ensures that the EV's battery is sufficiently charged to provide the necessary energy for the next trip. A similar model can be considered for BESSs but they can operate in all hours of day.

$$x_{e,t}^{\text{ch-EV}} + x_{e,t}^{\text{dch-EV}} \le 1, \quad \forall e\in\Omega_n^E, t\in\left[a_e^{\mathrm{EV}}, d_e^{\mathrm{EV}}\right] \quad (16)$$

$$x_{e,t}^{\text{ch-EV}} = 0, \ \ x_{e,t}^{\text{dch-EV}} = 0, \quad \forall e\in\Omega_n^E, t\in\Omega^T - \left[a_e^{\mathrm{EV}}, d_e^{\mathrm{EV}}\right] \quad (17)$$

$$0 \le p_{e,t}^{\text{ch/dch-EV}} \le x_{e,t}^{\text{ch/dch-EV}} \bar{P}_e^{\text{ch/dch-EV}}, \quad \forall e\in\Omega_n^E, t\in\Omega^T \quad (18)$$

$$SoC_{e,t}^{\mathrm{EV}} = SoC_{e,t}^{\text{EV-0}} + \left(\eta_e^{\text{ch-EV}} p_{e,t}^{\text{ch-EV}} - \frac{1}{\eta_e^{\text{dch-EV}}} p_{e,t}^{\text{dch-EV}}\right)\Delta t, \quad \forall e\in\Omega_n^E, t = a_e^{EV} \quad (19)$$

$$SoC_{e,t}^{EV} = SoC_{e,t-1}^{EV} + \left(\eta_e^{\text{ch-EV}} p_{e,t}^{\text{ch-EV}} - \frac{1}{\eta_e^{\text{dch-EV}}} p_{e,t}^{\text{dch-EV}}\right)\Delta t, \quad \forall e\in\Omega_n^E, a_e^{EV} < t \le d_e^{EV} \quad (20)$$

$$0 \le SoC_{e,t}^{EV} \le S_e^{\text{EV}}, \forall\, e \in \Omega_n^E, t \in \left[a_e^{\text{EV}}, d_e^{\text{EV}}\right] \tag{21}$$

$$SoC_{e,t=d_e^{\text{EV}}}^{\text{EV}} \ge E_e^{\text{trip-EV}}, \qquad \forall\, e \in \Omega_n^E \tag{22}$$

The flexible loads can be either interruptible or shiftable. Constraint (23) ensures that the utilized responsive demand remains within the allowable bounds, while (24) ensures that the minimum required energy is supplied after their response.

$$-\bar{P}_{f,t}^{\text{FL}} \le p_{f,t}^{\text{FL}} \le \bar{P}_{f,t}^{\text{FL}}, \qquad \forall\, f \in \Omega_n^F, t \in \Omega^T \tag{23}$$

$$\sum_{t\in\Omega^T}\left(P_{n,t}^{\text{Fixed}} + \sum_{f\in\Omega_n^F} p_{f,t}^{\text{FL}}\right).\Delta t \ge \underline{E}_n^{\text{L}}, \qquad \forall\, n \in \Omega^N, t \in \Omega^T \tag{24}$$

*2) Stage II: Local Flexibility Market*

The objective of this stage is to minimize load curtailments ($\Delta p_{n,t}^{\text{Fixed}}$) during disturbances and to fairly distribute the impact of the disturbances among different consumers by minimizing the difference between maximum and minimum pro-rated load curtailment, as expressed in (25).

$$\textbf{Min} \sum_{n\in\Omega^N}\sum_{t\in\Omega^T}\left(\Delta p_{n,t}^{\text{Fixed}} + \sum_{a\in\Omega_n^A}\sum_{a'\in\Omega_n^A}\left(\left(\overline{\frac{\Delta p_{a,t}^{\text{Fixed}}}{P_{a,t}^{\text{Fixed}}}}\right) - \left(\underline{\frac{\Delta p_{a',t}^{\text{Fixed}}}{P_{a',t}^{\text{Fixed}}}}\right)\right)\right) \tag{25}$$

When a disturbance occurs at different buses of the distribution network (the disturbance can either be an increase in load, $\Delta p_{n,t}^{\text{Dist}} \ge 0$, or a decrease in load, $\Delta p_{n,t}^{\text{Dist}} \le 0$), the flexibility of DERs is first optimally utilized in the form of upward and downward adjustments to minimize load shedding during disturbances. As a last resort, when load shedding becomes inevitable, consumers' demand is curtailed in a fair manner. Equation (26) ensures that the network power balance is maintained under abnormal conditions. It is worth noting that the network operational constraints (5) – (8) should be addressed in this stage for new adjusted power of DERs.

$$\begin{aligned}&\sum_{d\in\Omega_n^D}\left(UF_{d,t}^D - DF_{d,t}^D\right) + \Delta p_t^{\text{UG}}\Big|_{n\in\Omega^N,\text{PCC}} + \Delta p_{n,t}^{\text{Fixed}} + \sum_{m\in\Omega_n^N}\Delta p_{n,m,t}^{\text{F}} \\ &= \Delta p_{n,t}^{\text{Dist}}, \qquad \forall n\in\Omega^N, t\in\Omega^T\end{aligned} \tag{26}$$

It should be noted that the decision variables obtained from the first-stage problem are used as inputs for the second-stage problem. The flexibility of EVs in the real-time flexibility market is modeled through Equations (27) – (32). Constraint (27) specifies that EVs' flexibility can only be utilized when the EVs are in parking mode. Moreover, Constraint (28) ensures that EVs' upward and downward flexibility cannot be exploited simultaneously.

$$y_{e,t}^{\text{UF},E}, y_{e,t}^{\text{DF},E} = 0, \quad \forall e \in \Omega^E, \forall\, t \in \Omega^T - \left[a_e^{\text{EV}}, d_e^{\text{EV}}\right] \tag{27}$$

$$y_{e,t}^{\text{UF},E} + y_{e,t}^{\text{DF},E} \le 1, \quad \forall e \in \Omega^E, \forall\, t \in \left[a_e^{\text{EV}}, d_e^{\text{EV}}\right] \tag{28}$$

The optimal downward flexibility power of the EV is specified through Equations (29) and (30), which can either correspond to the allowable increase in charging power (if the EV is in charging mode at that time) or the allowable decrease in discharging power (if the EV is in discharging mode). Similarly, relations (31) and (32) specify the EV's upward flexibility power. The flexibility model for BESSs can be formulated in a manner similar to that of EVs.

$$0 \le DF_{e.t}^E \le y_{e.t}^{\text{DF},E}\left(\left(\bar{P}_e^{\text{ch-EV}} x_{e.t}^{\text{ch-EV}} - p_{e,t}^{\text{ch-EV}}\right) + p_{e,t}^{\text{dch-EV}}\right) \tag{29}$$

$$0 \le DF_{e.t}^E \le \left(\overline{SoC}_e^{\text{EV}} - SoC_{e,t}^{\text{EV}}\right) / \Delta t \tag{30}$$

$$0 \le UF_{e.t}^E \le y_{e.t}^{\text{UF},E}\left(\left(\bar{P}_e^{\text{dch-EV}} x_{e.t}^{\text{dch-EV}} - p_{e,t}^{\text{dch-EV}}\right) + p_{e,t}^{\text{ch-EV}}\right) \tag{31}$$

$$0 \le UF_{e,t}^E \le \left(SoC_{e,t}^{\text{EV}} - \underline{SoC}_e^{\text{EV}}\right) / \Delta t \tag{32}$$

The flexibility utilization model of flexible loads in local flexibility market are expressed in (33) – (35).

$$y_{f,t}^{\text{UF},F} + y_{f,t}^{\text{DF},F} \le 1, \ \forall f \in \Omega^F, t \in \Omega^T \tag{33}$$

$$0 \le UF_{f,t}^F \le y_{f,t}^{\text{UF},F}.p_{f,t}^{\text{FL}}, \qquad \forall f \in \Omega^F, t \in \Omega^T \tag{34}$$

$$0 \le DF_{f.t}^F \le y_{f,t}^{\text{DF},F}\left(\bar{P}_{n,t}^{\text{FL}} - p_{f,t}^{\text{FL}}\right), \qquad \forall f \in \Omega^F, t \in \Omega^T \tag{35}$$

The PVs only able to provide flexibility in downward direction by curtailing power as stated in (36).

$$0 \le DF_{p,t}^P \le p_{p,t}^{\text{PV}}, \qquad \forall p \in \Omega^P, t \in \Omega^T \tag{36}$$

Conventional DGs' flexibility utilization model is described in (37) – (41). The ramp rate constraints for DGs, as specified in (40) and (41), are included to limit changes in their output power within the flexibility market.

$$y_{c,t}^{\text{UF},C} + y_{c,t}^{\text{DF},C} \le 1, \qquad \forall c \in \Omega^C,\, t \in \Omega^T \tag{37}$$

$$0 \le UF_{c,t}^C \le y_{c,t}^{\text{UF},C}\left(\bar{P}_c^{\text{DG}} - p_{c,t}^{\text{DG}}\right), \qquad \forall c \in \Omega^C, t \in \Omega^T \tag{38}$$

$$0 \le DF_{c,t}^C \le y_{c,t}^{\text{DF},C}\left(p_{c,t}^{\text{DG}} - \underline{P}_c^{\text{DG}}\right), \qquad \forall c \in \Omega^C, t \in \Omega^T \tag{39}$$

$$0 \le UF_{c,t}^C \le RR_c^{\text{U-DG}}.\Delta t, \qquad \forall c\in\Omega^C, t\in\Omega^T \tag{40}$$

$$0 \le DF_{c,t}^C \le RR_c^{\text{D-DG}}.\Delta t, \qquad \forall c\in\Omega^C, t\in\Omega^T \tag{41}$$

## III. Numerical Results

To validate the proposed framework, it has been implemented on the modified IEEE 33-bus test distribution network [18]. It is assumed that DGs and BESSs penetration level is 30%, with 15% of the total loads considered flexible. Moreover, 10% of the actors in the network are assumed to own EVs. Detailed data on the DERs can be found in [4]. Furthermore, it is assumed that low-income actors are located at buses 2–5, 14–19, and 28–33, while medium-income actors are situated at buses 6–10 and 19–23. The remaining buses are assumed to host high-income actors.

The optimal management of DERs and the adjusted DLMP (during the peak price hour) in the proposed model, alongside the DLMP from the case where the energy burden index is ignored, is shown in Fig. 2. As observed, the adjusted DLMP is lower for low- and medium-income actors in comparison with high-income actors. Comparing the proposed method with the case where energy equity is not considered, the price adjustment in the proposed model results in a 0.75% reduction in social welfare. However, this aligns with the objective of energy equity, ensuring that low-income customers are not burdened with excessive costs to maximize overall social welfare.

When a 25% random load increase disturbance is applied to the network, the results associated with the second stage problem are summarized in Fig. 3. Comparing the proposed method with the case where equity is ignored and the objective is solely to minimize total load curtailments, the proposed model demonstrates a more distributed load curtailment across all buses. This results in a flatter violation in dispatched load across network buses, whereas in the latter case, significant dispatched load deviations occur at some nodes while other nodes experience no load curtailment. In the equity-based real-time flexibility market clearing framework, the total load curtailment is 3.55%. In contrast, when the objective is minimizing total load curtailment without equity considerations, the curtailment is slightly lower at 3.38%. This additional 5% load curtailment in the proposed model ensures a fair distribution of curtailments across all actors, considering their respective sizes, and upholds energy equity principles. Furthermore, in the absence of flexible resources, as depicted in Fig. 3, the system experiences a significantly higher load interruption of 5.93%. This highlights the critical role of flexible DERs in reducing curtailments in local communities.

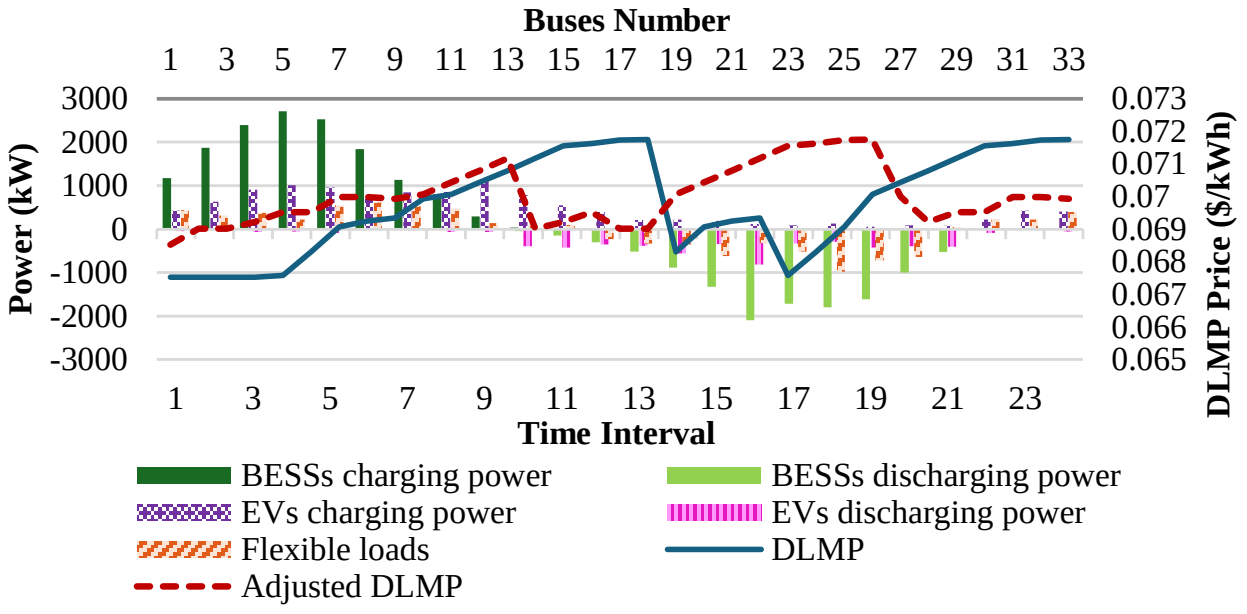

Figure 2. Optimal DERs management and DLMP signal in Stage 1.

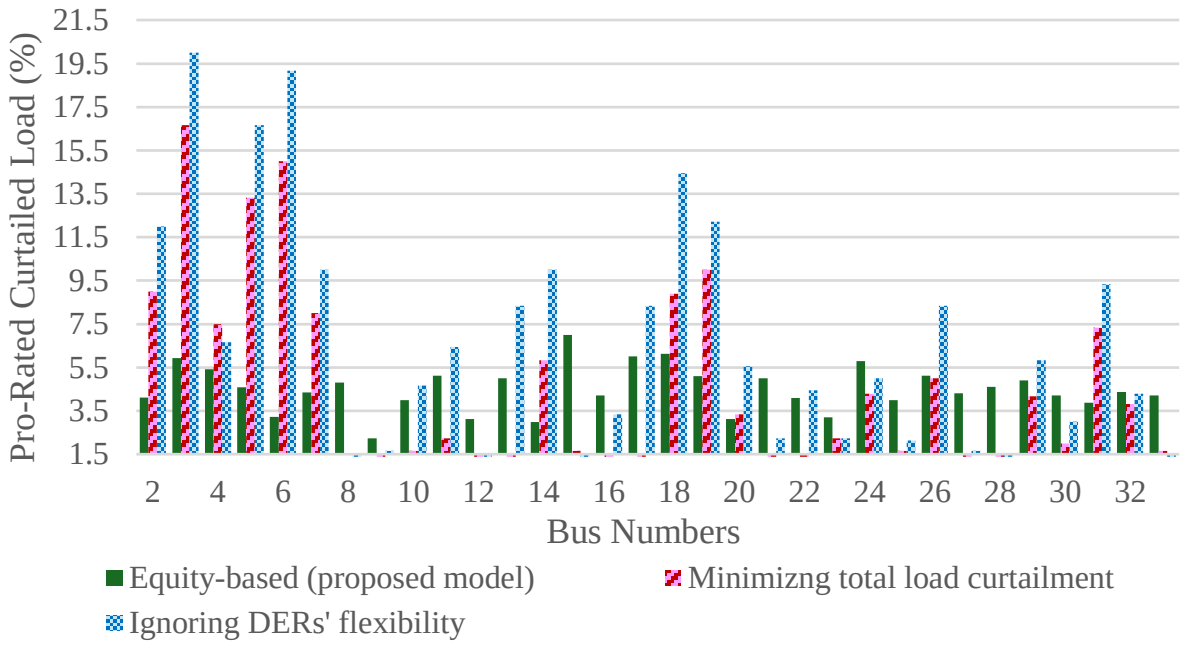

Figure 3. Load curtailment in case of disturbane occurrence.

## IV. Conclusion

This paper bridges the gap between social equity and the technical and economic aspects of local electricity markets. A two-stage framework is proposed, where the first stage focuses on the optimal operation of actors' DERs in a day-ahead local energy market. In this stage, the DLMP signal is adjusted to ensure fair energy pricing within the local community by considering the energy burden index. In the second stage, the flexibility of DERs is leveraged in the real-time local flexibility market to maintain energy equity in the event of network disturbances, minimizing the pro-rated load curtailment among consumers and prosumers. The results from the first stage indicate a reduction in social welfare, but this is offset by the distribution of energy at lower prices for low income actors. Furthermore, during abnormal conditions, the flexibility potential of DERs is optimally utilized, ensuring that the minimum required load curtailment is fairly distributed across different consumers, while accounting for their initial energy dispatch from the first stage.

## References

[1] O. W. Johnson, J. Y.-C. Han, A.-L. Knight, S. Mortensen, M. T. Aung, M. Boyland, and B. P. Resurreccion, "Intersectionality and energy transitions: A review of gender, social equity and low-carbon energy," *Energy Research & Social Science*, vol. 70, p. 101774, 2020.

[2] H. Chen, L. Fu, L. Bai, T. Jiang, Y. Xue, R. Zhang, B. Chowdhury, J. Stekli, and X. Li, "Distribution market-clearing and pricing considering coordination of DSOs and ISO: An epec approach," *IEEE Transactions on Smart Grid*, vol. 12, no. 4, pp. 3150–3162, 2021.

[3] Y. Cai, T. Huang, E. Bompard, Y. Cao, and Y. Li, "Self-sustainable community of electricity prosumers in the emerging distribution system," *IEEE Transactions on Smart Grid*, vol. 8, no. 5, pp. 2207–2216, 2016.

[4] N. Pourghaderi, M. Fotuhi-Firuzabad, M. Moeini-Aghtaie, M. Kabirifar, and P. Dehghanian, "A local flexibility market framework for exploiting DERs' flexibility capabilities by a technical virtual power plant," *IET Renewable Power Generation*, vol. 17, no. 3, pp. 681–695, 2023.

[5] Y. He, Q. Chen, J. Yang, Y. Cai, and X. Wang, "A multi-block admm based approach for distribution market clearing with distribution locational marginal price," *International Journal of Electrical Power & Energy Systems*, vol. 128, p. 106635, 2021.

[6] N. Pourghaderi, M. Fotuhi-Firuzabad, M. Moeini-Aghtaie, M. Kabirifar, and M. Lehtonen, "Exploiting DERs' flexibility provision in distribution and transmission systems interface," *IEEE Transactions on Power Systems*, vol. 38, no. 2, pp. 1963–1977, 2022.

[7] A. Chapman, Y. Shigetomi, H. Ohno, B. McLellan, and A. Shinozaki, "Evaluating the global impact of low-carbon energy transitions on social equity," *Environmental Innovation and Societal Transitions*, vol. 40, pp. 332–347, 2021.

[8] V. C. Johnson, S. Hall, J. Barton, D. Emanuel-Yusuf, N. Longhurst, A. O'Grady, E. Robertson, F. Sherry-Brennan, and E. Robinson, "Community energy and equity: The distributional implications of a transition to a decentralised electricity system," *People, Place and Policy*, vol. 8, no. 3, pp. 149–167, 2014.

[9] B. K. Sovacool, R. J. Heffron, D. McCauley, and A. Goldthau, "Energy decisions reframed as justice and ethical concerns," *Nature Energy*, vol. 1, no. 5, pp. 1–6, 2016.

[10] B. L. Tarufelli and S. R. Bender, "Equity in transactive energy systems," *in 2022 IEEE PES Transactive Energy Systems Conference (TESC).* IEEE, 2022, pp. 1–5.

[11] C. A. Adams and S. Bell, "Local energy generation projects: assessing equity and risks," *Local Environment*, vol. 20, no. 12, pp. 1473–1488, 2015.

[12] L. Rodriguez-Garcia, A. Hassan, and M. Parvania, "Equity-aware power distribution system restoration," *IET Generation, Transmission & Distribution*, vol. 18, no. 2, pp. 401–412, 2024.

[13] M. Nazari-Heris, A. Loni, S. Asadi, and B. Mohammadi-ivatloo, "Toward social equity access and mobile charging stations for electric vehicles: A case study in Los Angeles," *Applied Energy*, vol. 311, p. 118704, 2022.

[14] Q. Zhang and F. Li, "Securing energy equity with multilayer market clearing," *IEEE Transactions on Energy Markets*, Policy and Regulation, 2024.

[15] Z. Li, C. S. Lai, X. Xu, Z. Zhao, and L. L. Lai, "Electricity trading based on distribution locational marginal price," *International Journal of Electrical Power & Energy Systems*, vol. 124, p. 106322, 2021.

[16] S. Cong, D. Nock, Y. L. Qiu, and B. Xing, "Unveiling hidden energy poverty using the energy equity gap," *Nature communications*, vol. 13, no. 1, p. 2456, 2022.

[17] C. B. Domenech, J. Naughton, S. Riaz, and P. Mancarella, "Towards distributed energy markets: Accurate and intuitive DLMP decomposition," *IEEE Transactions on Energy Markets, Policy and Regulation*, 2024.

[18] D. Feroldi and P. Rullo, "Optimal operation for the IEEE 33 bus benchmark test system with energy storage," *2021 IEEE URUCON*, Montevideo, Uruguay, 2021, pp. 1-5.